\documentclass[runningheads]{llncs}
\usepackage[T1]{fontenc}
\usepackage{graphicx}  

\usepackage{amsmath, amsfonts, amssymb, amscd, enumerate}
\usepackage{verbatim}

\usepackage{amsmath}
\usepackage{amssymb}
\usepackage{amsfonts}
\usepackage{hyperref}

\def\vp{\varphi}
\def\g{\gamma}

\def\O{\Omega}

\DeclareFontFamily{OT1}{msb}{}{}
\DeclareFontShape{OT1}{msb}{m}{n}
{  <5> <6> <7> <8> <9> <10> gen * msbm
	<10.95><12><14.4><17.28><20.74><24.88>msbm10}{}
\DeclareMathAlphabet{\bubble}{OT1}{msb}{m}{n}

\def\bR{{\mathbb R}}
\def\bZ{{\mathbb Z}}

\def\bH{{\mathbb H}}

\newfont{\goth}{eufm10 scaled \magstep1}

\def\Sac{{\mathrm{Sac}}}
\newfont{\mcal}{eusm10 scaled \magstep1}

\def\bp{\bar{\partial}}

\def\span{\mathrn{span\;}}

\def\Sac{\mathrm{Sac\;}}

\def\Ad{\mathrm{Ad}}

\def\Sac{\mathrm{Sac\;}}

\def\span{ \mathrm span \;}
\newtheorem{Th}{Theorem}
\newtheorem{Prop}{Proposition}
\newtheorem{Cor}{Corollary}
\newtheorem{Lem}{Lemma}
\newtheorem{Def}{Definition}
\newtheorem{Rem}{Remark}
\def\bt{\begin{Th}}
	\def\et{\end{Th}}
\def\bp{\begin{Prop}}
	\def\ep{\end{Prop}}
\def\bc{\begin{Cor}}
	\def\ec{\end{Cor}}
\def\bl{\begin{Lem}}
	\def\el{\end{Lem}}
\def\bd{\begin{Def}}
	\def\ed{\end{Def}}
\def\br{\begin{Rem}}
	\def\er{\end{Rem}}
\def\pf{\noindent{\it Proof. }}
\def\qed{\hspace{2ex} \hfill $\square $ \par \medskip}

\def\be{\begin{equation}}
	\def\ee{\end{equation}}
\def\arr{\begin{array}{rlll}}
	\def\ea{\end{array}}
\def\bea{\begin{eqnarray}}
	\def\eea{\end{eqnarray}}
\def\bean{\begin{eqnarray*}}
	\def\eean{\end{eqnarray*}}

\usepackage{textcomp}

\begin{document}
	\title{Geometry of saccades and saccadic cycles  \thanks{D. V. Alekseevsky was supported
			by the Grant "Basis-foundation (Leader)" 22-7-1-34-1. I.M. Shirokov was supported
			by the Grant "Basis-foundation (Leader)" 22-7-1-34-1 and Ministry of Science and Higher Education of
			the Russian Federation, agreement no. 075-15-2022-289.}}
	%
	%
	\author{D. V. Alekseevsky \inst{1,2}\orcidID{0000-0002-6622-7975} \and
		I.M. Shirokov \inst{3}\orcidID{0009-0005-1538-2712}}
	\authorrunning{D. V. Alekseevsky, I.M. Shirokov}
	%
	\institute{Institute  for  Information   Transmission Problems, B. Karetnuj per., 19, Moscow, 127051, Russia \and
		University of Hradec Kr\'alov\'e, Faculty of Science, Rokitansk\'eho 62, 500~03 Hradec Kr\'alov\'e,  Czech Republic \\
		\email{dalekseevsky@iitp.ru}\\
		\and
		St. Petersburg Department of Steklov Mathematical Institute Fontanka, 27, 191023, St. Petersburg, Russia \\
		\email{shirokov.im@phystech.edu}}
	\maketitle              
	\begin{abstract}
		
		The paper is devoted to the development of the differential geometry of saccades and saccadic  cycles. We recall an interpretation of Donder's and Listing's law in terms of the Hopf fibration of the $3$-sphere over the $2$-sphere. In particular, the configuration space of the eye ball (when the head is fixed) is the 2-dimensional hemisphere $S^+_L$, which is called Listing's hemisphere. We give  three  characterizations  of saccades: as geodesic  segment $ab$ in the Listing’s   hemisphere, as the   gaze   curve and  as  a  piecewise geodesic  curve   of the orthogonal group. We study   the  geometry  of  saccadic  cycle,  which is  represented  by a geodesic polygon  in the  Listing hemisphere, and give  necessary and  sufficient  conditions, when  a system of lines through  the center of  eye ball is the system of axes of rotation for saccades of the  saccadic  cycle, described in terms of world  coordinates and  retinotopic  coordinates. This gives an  approach  to  the  study the  visual stability problem.
		
		\keywords{Donder's and Listing's law \and quaternions \and Hopf bundle \and fixation
			eyes movements \and drift \and micrsasccades \and remapping \and shift of receptive fields \and
			neurogeometry}
	\end{abstract}
	\section{Introduction}
	
	Our eyes continually move.  Even while we fix our gaze on an object, they participate in  fixational eye movements -- tremor, drift and microsaccades (see Fig. 1.).  The  classical experiments by A. Yarbus  show  that compensation of  eye movements leads  to loss of vision in 2-3 sec. The  information about saccades is coded into control commands of the oculomotor system which governs muscles contraction.
	A copy of these command (efference  copy  or corollary discharge) is send to the frontal eye field of   the frontal cortex \cite{W}, where it meets the statistical information  about   retina images, described  in coordinates  associated to the  end  gaze  direction of the  previous saccade.
	
	\begin{figure}[h!]
		\centering
		\includegraphics[scale = 0.6]{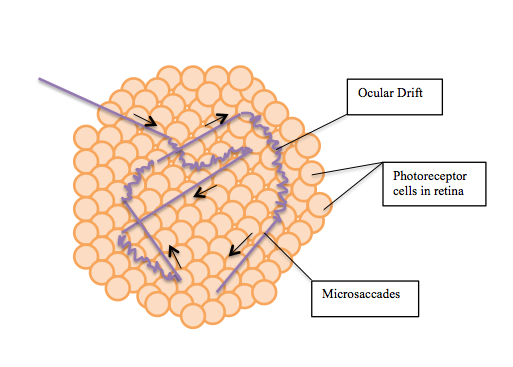}
		\caption{Fixational eye movements (tremor is not shown). From Wikipedia, The Free Encyclopedia, 2022. Available at: \href{https://en.wikipedia.org/w/index.php?title=Fixation_(visual)&oldid=1136172595}{Visual fixation}} \label{Fig. 1.}
	\end{figure}

	The vision is the  result of  interaction of these two  types of information. The  time synchronization   of oculomotor   information about   eye  rotations   and  retinal information is very important  for the  correct decoding.
	
	In  \cite{A} a  new interpretation of  the Donder's and Listing’s laws in terms of  the Hopf bundle  had been  given and used for   geometrical  descriptions of   saccades and drift.
	This  paper,  which is  a  continuation of    \cite{A},  is devoted to development of   the   differential geometry of   saccades and saccadic  cycles. We    give  three  characterizations  of saccades --  as    geodesic  segment $ab$ in the Listing’s   hemisphere  (which is the configuration space of the  eye), as the   gaze   curve
	$Sac(AB)  = \chi(a)\chi(b)  \subset  S^2_E\cap \span(A,B,-i)$,
	where $\chi : S^3 \to S^2$ is the   Hopf  projection, and  as  a  piecewise geodesic  curve   of the orthogonal group.
	Then we study   the    geometry  of  saccadic  cycle  which is  represented  by a   geodesic polygon  in the  Listing hemisphere.
	We  give  necessary and  sufficient  condition, when  a system of lines
	$([\O_1],\cdots, [\O_n])$   , where  $\O_{\ell} \in S^2_E$,  is  the system of    axes  for saccades  of the  saccadic  cycle,  described  in terms of world coordinates and retinotopic coordinates.
	
	This  clarifies   the  relation between oculomotor  information  about  saccadic  rotation the    gaze   direction    and hence  the   eye position   and  shows  that  all  visual information  may  be  transformed into world-centered coordinates, which gives an approach  to  the  study of the  visual stability problem.
	
	\section{The Hopf bundle and the Listing's section}
	
	Let
	$$\bH = \bR^4 = \span(1,i,j,k) = \bR 1 + E^3, \, E^3 = \mathrm{Im}\bH=\span(i,j,k)$$ be  the  algebra of quaternions with   the  standard  Euclidean  metric \\ $|q|^2 = <q, q> = q \bar{q}$, where  $\bar{q} = q_R - q_E $ is  the  conjugation.
	
	The  sphere   $\bH_1=S^3$  of unit  quaternions is the  Lie  group   and the  adjoint  action
	$$ \Ad:  \bH_1  \to SO(E^3),\,a \mapsto \Ad_a,\, \Ad_a x = axa^{-1}=ax\bar{a}$$
	is   the universal   $\bZ_2$-covering.\\
	
	The orbit  $\Ad_{\bH_1}i = S^2_E \simeq \bH_1/SO(2)$  is the unit sphere and $\chi: S^3 \to S^2_E$ is the principal bundle.
	
	\bd
	The   Hopf bundle  is  the  natural $\bH_1$-equivariant map
	$$\chi : S^3 = \bH_1  \to S^2_E,\,  a \mapsto A= \chi(a) := \Ad_a i = aia^{-1}.   $$
	\ed
	
	Denote  by $S^2_L = S^3 \cap \span(1,j,k)= S^3 \cap i^{\perp}$ the  equatorial 2 -sphere   of  $S^3$ w.r.t.  the poles $\pm i$ and  by
	$  S^+_L = \{  a = a_0 1 + a_2 j + a_3 k \in S^2_L, \,  a_0 >0\}$
	the   north  hemisphere.  It is called  the Listing  hemisphere  \cite{A}  and its  boundary
	$S^1_L$ is called  the Listing's   circle (see Fig. 2.).
	
	\bp The (restricted  to $S^+_L$) Hopf   map
	$$\chi: S^+_L  \to  S^2_E,\,      a \mapsto    A =  \chi(a) =  \Ad_a i  =  aia^{-1}$$
	is a diffeomorphism   of $S^+_L$ onto the punctured   sphere $\tilde{S}^2_E  = S^2_E \setminus \{-i\}$ and $\chi(S^1_L) =\{-i\}.$
	\ep
	
	\bd 
	i) The  map $\sigma=\chi^{-1} :\tilde{S}^2_E \to S^+_L $ is called  the Listing's section.\\
	ii) The  frame $ f^1 = (i,j,k)$ is called the  standard   and   the  frame
	$f^a : = \Ad_a f^1 = (aia^{-1}, aja^{-1}, aka^{-1}) $ is called admissible.
	\ed
	
	\bp \cite{A} i)Any two points  $a,b \in S^+_L$   determine  a unique (oriented)  geodesic of the Listing's sphere with canonical parametrization
	$$ \g_{a,b} = \g_{p,m}(t) =\cos t\,p + \sin t \,m = e^{tv}p $$
	where $p ={\cos \theta}\,j +\sin{\theta}\,k $ is the intersection of $\g_{a,b}$ with  $S^1_L$, $ m = e^{\psi q}= \cos \psi + \sin \psi q,\, q=ip$ is the  top point of the  geodesic  and $v  = m\bar{p} = \sin \psi \,i - \cos \psi p \in S^2_E$.\\
	ii)  If $m=1$, then $\g_{a,b} =\g_{p,1} = e^{tp}, \, p \in S^1_L$ is 1-parameter  subgroup. \\
	iii) The group $\Ad_{e^{tp}} = R^{2t}_{p}$ is the group  of rotation w.r.t.  the axis $[p] := \bR p$ and the Hopf image $\chi(\g_{p,m})= \Ad_{e^{tv}}(-i) = R^{2t}_v(-i)$ is the gaze  circle  $S^1_{a,b}=S^1_{p,m} = S^2_E \cap \Pi(A,B,-i)$,  where  $A= \chi(a),\, B = \chi(b)$ and $\Pi(A,B, -i) = -i +\span(i+A,i+B)$.
	\ep
	
	\begin{figure}[h!]
		\centering
		\includegraphics[scale = 0.3]{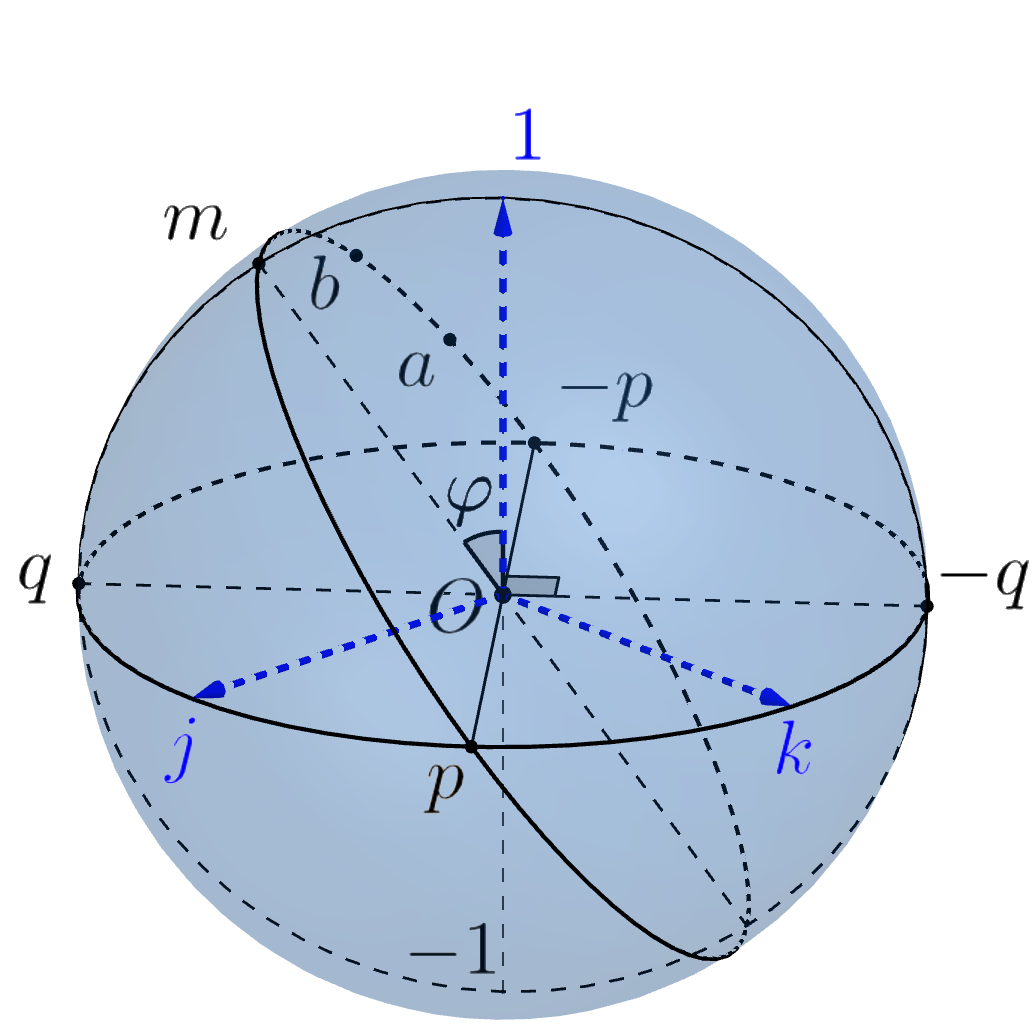}
		\caption{The Listing's sphere.} \label{Fig. 2.}
	\end{figure}
	
	\begin{figure}[h!]
		\centering
		\includegraphics[scale = 0.3]{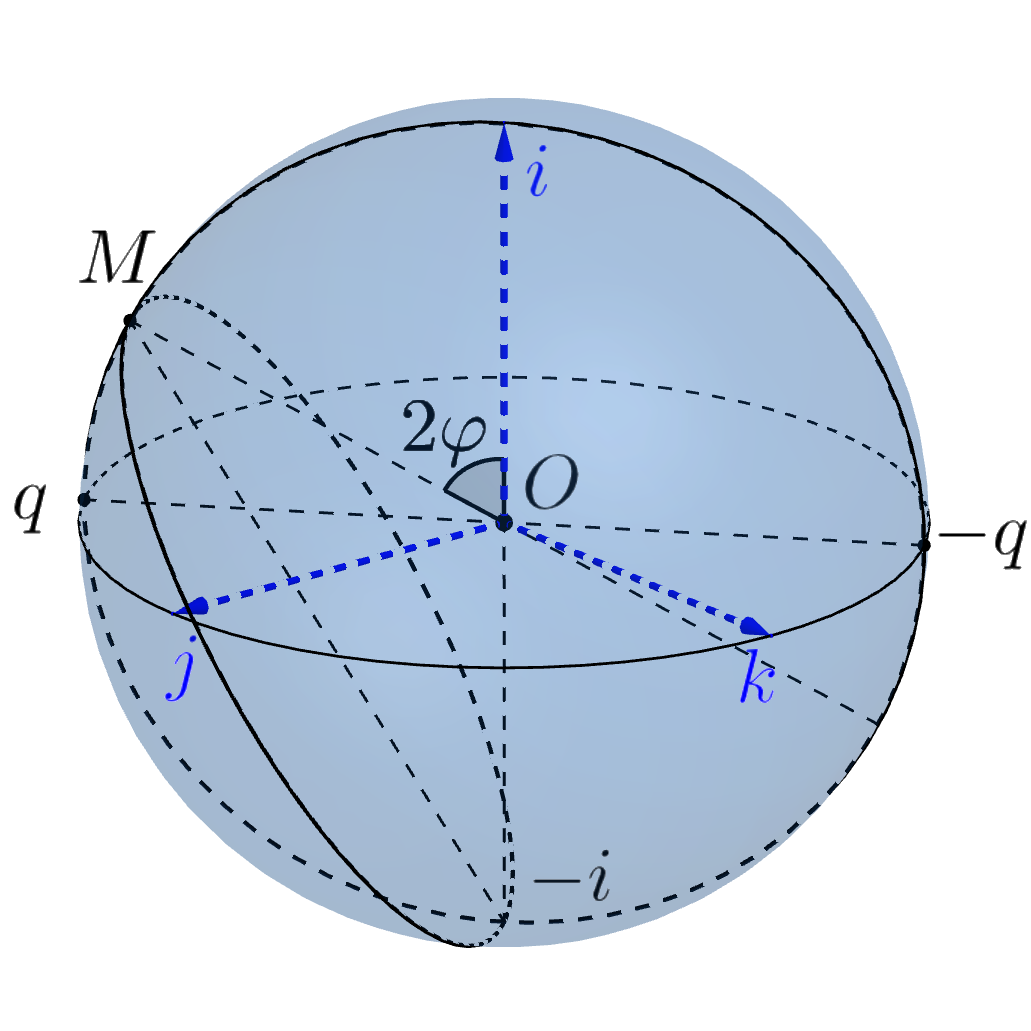}
		\caption{The eye sphere.} \label{Fig. 3.}
	\end{figure}
	
	\section{Donder's  and Lising's  laws}
	We  consider   the  sphere  $S^2_E \subset E^3$ in the Euclidean space  $E^3$ as a model of the  eye  sphere (see Fig. 3.), i.e. the boundary of the  eye ball in  the  primary position,   defined  by the  primary   frame $f^1 = (i,j,k)$. Here $i$ indicates  the  primary position of the gaze (the frontal direction),  $j$  is the lateral direction from  right  to left and   $k$ is the  vertical direction  up.
	
	\emph{Donder's law}. If the head is fixed,  the  configuration space  of the eye is  2-dimensional and the position  $A \in S^2_E $ of  the  gaze determines  the position   of the eye.\\
	The Listing's  law  describes the   configuration space.
	
	\emph{Listing's law}. The  configuration space of the  eye   is the Listing's hemisphere.  The  eye position   associated to a   gaze  $A \in \tilde{S}^2_E$ is  given by
	$$f(A)= f^{\sigma(A)}:=  \Ad_{\sigma(A)}f^1.$$

	\section{Saccades, three  descriptions}
	
	The  saccade $\Sac(A,B)$  from a gaze direction  $A \in \tilde{S}^2_E$  to the  gaze direction $B$  is  described  by the   geodesic  segment  $ab = \sigma(A) \sigma(B)$   and   the   corresponding gaze curve  $ AB=\chi(ab)$, which  is the segment  of the gaze circle  $S^1_{a,b}$ with the  endpoints  $A$ and $B$.
	
	Let    $a,b \in S^+_L $ be two  eye positions   and   $A = \chi(a), B = \chi (b) \in S^2_E$  the  associated   gaze directions.  It  defines  a unique    saccade  $\Sac(A,B)$ from  the  eye position  $a$ to the  position $b$. 
	
	We  have   three  descriptions  of the saccade $\Sac(A,B)$:\\
	i) As the  evolution of  the   eye position  in the configuration space  $S^{+}_L$  that  is represented  by a  segment  $ab$ of a  geodesic $\gamma_{a,b} \subset S^{+}_L$ of the Listing  hemisphere. \\
	ii) As  the   evolution of  the gaze in the  gaze  sphere  $\tilde{S}^2_E$,  described  by the  gaze curve $AB=\chi(ab)$.
	It  is   the  segment of the  gaze circle
	$S^1(A,B)  =    S^2_E \cap  \Pi(A, B, -i)$   which is the   section of the  sphere $S^2_E $ by the   plane  $\Pi(A,B, -i)  = -i + \span(A+i, B +  i)$.\\
	iii) As the  curve  $R^{t\vp}_{\O},\,   t \in [0,1]$ in the orthogonal group,  which  is  a segment   of the  one-parameter group of rotations  around   the   axis $[\O] = \bR \O$.
	
	To  relate  these three characterisations of saccades, we  introduce the  following    definition.
	
	\bd  Let   $A \in \tilde{S}^2_E$  be a unit  vector. Then  the  oriented 2-dimensional subspace $L_A = N_A^{\perp}$ with the  positive  unit normal $N_A  = \frac{i+A}{|i+A|}$  is called   $A$-Listing  plane   and  $N_A$  is called  $A$-Listing's  normal.
	\ed
	\bp  i) The   axis $[\O] = \bR \O$ of   rotation of  any saccade    with  initial  gaze direction   $A \in \tilde{S}^2_E$  is  orthogonal  to the $A$-Listing's  normal $N_A$ and  belongs to the  $A$-Listing's plane $L_A = N_A^{\perp}$. \\
	ii) \textbf{Law of half angle}: the  ( unique) saccade  $\Sac(A,B)$ from the gaze direction $A$ to the gaze direction $B$  is obtained  by  the  rotation  $R_{\Omega}^{\vp}$  around  the   axis
	$[\Omega] =  \bR( N_A  \times N_B) =L_A \cap L_B $
	by the  angle $\vp =2 \psi$  where   $\psi=\angle (N_A, N_B)$ is the  angle between  the  normal vectors $N_A,\,N_B$ of the A-Listing's plane
	$L_A$  and the $B$-Listing's  plane  $L_B$.\\
	iii) Let $Sac(A_0, A_1),\,   Sac(A_1, A_2)$ be two consecutive saccades with distinct axes of rotation
	$\O_1, \, \O_2$.  Then  the  $A_1$-Listing's plane   is given by
	$$L_{A_1} =\span(\O_1, \O_2)$$
	and
	the $A_1$-Listing's normal is given by:
	$$ N_{A_1} = \frac{i^{*}}{|i^{*}|}, $$
	where $i^{*}$ stands for the first dual vector in the dual basis to the basis $(i, \Omega_1, \Omega_2)$.
	
	\ep
	
	\pf  i) Let $B \in \tilde{S}^2_E \setminus \{A\}$. Then during the saccade $Sac(A,B)$ the gaze belongs to the circle $\Pi(A, B, -i) \cap S^2_E$. Hence the axis of rotation of $Sac(A,B)$ is orthogonal to the plane $\Pi(A, B, -i)$ and in particular orthogonal to the vector $N_A$. \\
	
	ii) Indeed,   denote    by   $O'$   the center   of the saccadic   circle   $S^1_{A,B}=  \Pi(A, B, -i) \cap S^2_{E}$. Then the   rotation angle  $\vp = \angle(O'A, O'B)$ is the central angle   associated   to the
	inscribed angle
	$$\psi =\angle ((-i)A, (-i)B)= \angle (N_A, N_B)  .$$
	Hence $\psi = \frac12 \vp.$\\
	
	iii) First note that $\O_1, \O_2$ and $i$ are linear independent. We have
	
	$$i^* = \frac{[\O_1, \O_2]}{(i, [\O_1, \O_2])}$$
	
	Thus $\mathbb{R}i^* = \mathbb{R}N_{A_1}$ and $(i^*,i) = 1 > 0$. So $N_{A_1} = \frac{i^{*}}{|i^{*}|}$.
	
	\qed
	
	\section{Saccadic  $n$-cycles}
	
	We define  a \emph{saccadic  $n$-cycle}  as  a geodesic  polygon $Sac(a_1 a_2 \cdots  a_{n+1}) \subset S^+_L, a_1 = a_{n+1}$ and $a_k \neq a_{k+1}$ with the natural parametrization, which is  identified  with the   gaze curve
	$$ \chi(Sac(a_1 a_2 \cdots  a_{n+1})=
	Sac(A_1, ..., A_n, A_{n+1}).
	$$
	
	In terms of the    rotations, the saccadic curve is described as the orbit
	$R^t  A_1$   of
	a piecewise  geodesic curve  $ R^t,\, t \in [0, T]$  in the orthogonal group  $SO(3)$
	such  that
	$$R^{T_j}A_1 = A_j,\, R^{T_j+t}A_1 = R^t_{\O_j}A_j,\,  0\leq t \leq T_{j+1}- T_j, \,$$
	$$ 0<T_1 < \cdots < T_{n+1} =T.$$
	We call  the vectors $\Omega_j$ which determine   the  axes  of  rotation $[\O_j] = \bR \O_j$  of  the  saccades  $Sac(A_j, A_{j+1})$ the  axes   of the $n$-cycle.
	
	\subsection{Saccadic $n$-cycles via systems of axes  expressed at the primary frame}


	Note   that  the   gaze  directions $A_{j}, A_{j+1}$   of   the   saccade $S_j = Sac(A_{j}, A_{j+1})$  determine the  saccade rotation $R^{\vp}_{\O_j}$     as follows\\
	$$ \O_j \equiv  N_{A_{j}} \times N_{A_{j+1}}, \,\,
	\vp_j = 2\angle (N_{A_{j}}, N_{A_{j+1}}) $$ 
	
	(Where "$\equiv$"  means equality up to sign). Now   we  consider  the inverse  problem:  when a system  $(\O_1, \cdots, \O_n)$  of  unit  vectors determines  a  saccadic  $n$-cycle   and  how  to describe  the  gaze directions   $(A_1, \cdots, A_{n+1} = A_1)$   and the  angles of  rotations of  the $n$-cycle.

	We    show  that it can be  done   under  a mild necessary and  sufficient  condition     and  give   an explicit  formula for  the  system of gaze  directions   $(A_1, \cdots, A_{n+1})$.
	\bt
	Let   $(\O_1, \cdots, \O_n)$ be a system of  $n\geq 3$ unit  vectors in $E^3 $.  Then  there exists   a (unique) saccadic $n$-cycle
	$$  Sac(A_1, A_2, \cdots , A_{n + 1}),\, A_{n+1} = A_1$$
	with  axes  of  rotation
	$$[\O_1], \cdots, [\O_n]$$
	if and only if
	\begin{enumerate}
		\item All triples $(\O_1, \O_2, i), \cdots, (\O_n, \O_1, i)$ are linear independent.
		
		\item All triples $(\O_1, \O_{2}, \O_{3}), \cdots, (\O_{n-1}, \O_{n}, \O_{1}), (\O_n, \O_{1}, \O_{2})$ are linear independent.
	\end{enumerate}
	
	If $span(\O_{n}, \O_1) = span(j,k)$ then $A_1 = A_{n+1} = i$.
	\et
	\pf
	Necessity follows from Proposition 1 and definition of saccadic cycle. The scheme of proof of sufficiency is as follows. Planes $$\Pi_1 := span(\O_n, \O_1), \Pi_2 := span(\O_1, \O_2), \cdots, \Pi_n := span(\O_{n-1}, \O_{n})$$ define points $a_1, \cdots, a_n$ of configuration space $S^+_L$ such that $\Pi_k$ is $\chi(a_k)$-Listing plane. It remains to check that $Sac(a_1, ..., a_n, a_{n+1} = a_1)$ is a desired cycle.
	
	\qed
	\subsection{Transition  functions  between   primary    frame and the moving  frame}
	
	
	In this  subsection  we  show  how coordinates of Listing's normal $N_A$ w.r.t the primary frame $f^1$ and "moving" frame $f^{\sigma(A)}$ are related.
	
	\bl \label{crucial lemma}
	Let $a \in S^+_L$ and $A = \chi(a)$ be the gaze direction.
	Then coordinates of $A$-Listing's normal $N_A$ w.r.t. the primary frame $f^1$ and "moving" frame $f^a$ related as follows:
	
	$$N_A = N_1 i + N_2j + N_3 k = N_1 i^a - N_2j^a - N_3 k^a$$
	\el
	
	\pf
	We consider the saccade $Sac(1,a) = Sac(i,A)$. By the half-angle law (Proposition 3) we have $N_A = a^{1/2}ia^{-1/2}$. We  may write $a = e^{\varphi p}$ for some $p = \cos \theta j + \sin \theta k$ and $\varphi \in [0,\pi/2)$, then w.r.t. the primary frame $f^1$, the vector $N_A$ is obtained from $i$ by rotation around $[p]$ on $\varphi/2$: $N_A =   R^{\varphi/2}_{[p]}  i$, whereas w.r.t. the "moving" frame $f^a$, the vector the vector $N_A$ is obtained from $i^a$ by rotation around $[p^a] = [\cos \theta j^a + \sin \theta k^a]$ on $-\varphi/2$: $N_A   =  R^{-\varphi/2 }_{[p^a]} i^a$.
	\qed
	
	\subsection{Saccadic $n$-cycles via systems of axes fixed at the moving frame}
	\bd
	We   associate  with  a vector $\O\in S^2_E$
	the  map
	$$\sigma_{\Omega}: S^+_L \to S^2_E, , a \mapsto \O^a = \Ad_a\O = a \O a^{-1} $$
	We  call   parameterized by $a \in S^+_L$ vector $\O^a$ the moving vector.
	\ed
	
	The coordinates $\O^a_{f^a}$ of  the vector $\O^a$ in  the   frame $f^a$  does not depend on $a$  and  coincide with  the  coordinates of  vector $\O$ w.r.t. the  primary frame.
	Note that according to the Listing's law
	the  eye  can rotate  around axis $[\O^a]$ if
	$\O^a$ belongs to the  $\chi(a)$-Listing's  plane. For  example,   $[\O^1]$  is  the   axis of  rotation only if it  belongs to  $\chi(1)=i$-Listing's   plane, which is $\span(j,k)$.
	
	Given a system  $(\O_1, \cdots, \O_n)$ of vectors from $S^2_E$, our aim  is to   describe    when
	there exists a saccadic $n$-cycle $Sac(a_1, ..., a_n, a_{n+1})$   with axes of  rotation
	$$   [\O_1^{a_1}], \cdots, [\O_n^{a_n}].                       $$
	
	We  need  the  following

	\bp  Assume the vectors  $\O_1,\O_2 \in S^2_E$    span    the plane  $\Pi = \span(\O_1, \O_2)$  which does not contains  $i$. Then there exists unique $a \in S^+_L$, hence gaze direction $A = \chi(a)$, such that $\Pi^a:=\Ad_a \Pi =  \span(\O_1^a, \O_2^a)$ coincides with  $A$-Listing plane $L_A$.
	\ep
	\pf
	Proposition 3   implies  that   there  is unique $b \in S^+_L$ such that $\Pi=\Ad_{b^{1/2}}L_i = L_{b}.$ We have to find  $a \in S^+_L$  such that
	$$\Pi^a = \Ad_{ab^{1/2}}L_i =L_a= \Ad_{a^{1/2}}L_i $$
	The solution is  $a = b^{-1}$.
	\qed
	\bt
	The   system  $(\O_1, \cdots, \O_n)$  of  vectors from $S^2_E$   defines  the system
	of   axes of rotation
	$$   ([\O_1^{a_1}], \cdots, [\O_n^{a_n}])                       $$
	for a  (unique) saccadic $n$-cycle
	$$ Sac(a_1, \cdots, a_{n+1}), \, a_1 = a_{n+1}$$
	if and only if it satisfies the  conditions of  Theorem 1.
	If $span(\O_{n}, \O_1) = span(j,k)$ then $a_1 = a_{n+1} = 1$.
	\et
	
	\pf
	
	We prove only sufficiency. Let $\O_{n+1}^a := \O_1^a$. Proposition 4 implies that for each pair $\O_k, \O_{k+1}, k \in \{1, ..., n\}$ there exists $a_k \in S^+_L$, such that $$\Pi_k^{a_k} := span(\O_k^{a_k}, \O_{k+1}^{a_{k}}) = L_{a_k} \eqno(1)$$

	Let $a_{n+1} := a_1$. Proof of Proposition 4 implies that $\Pi_k^{a_k} = S_{L_i}(\Pi_k^{1}) $, where
	$$ S_{L_i}: \xi_1 i + \xi_2 j + \xi_3 k \mapsto -\xi_1 i + \xi_2 j + \xi_3 k$$
	is  the  symmetry w.r.t.the Listing's   plane $L_i$. Hence we have
	$$\O_1^{a_n} = \O_1^{a_1}, \: \: \O_k^{a_{k-1}} = \O_k^{a_{k}}, k \in \{2, ..., n\} \eqno(2)$$
	
	We claim that $Sac(a_1, ..., a_n, a_{n+1})$ is a desired cycle. Indeed $a_k \neq a_{k+1}$ for $k \in \{1, ..., n\}$ due to the assumptions on $\O_1, ..., \O_n$, the geodesic $a_ka_{k+1}$ corresponds to rotation around $[\O_k^{a_k}]$ due to (2) for each $k \in \{1, ..., n\}$; and $[\O_k^{a_k}] \subset L_{a_k}$ for each $k \in \{1, ..., n\}$ due to (1).
	\qed
	
	\begin{Cor}
		Let $Sac(a_1, ..., a_{n+1})$ be a saccadic cycle described in terms of Theorem 4 and $[\O_m] ,\, m=1, \cdots, n$   be   axes  of  rotation.  
		Then the   coordinates of any   vector $\O := \O^{a_m}_m$  w.r.t. the “moving“ frame
		$f^{a_m}$ and the primary frame $f^1$ are:
		
		$$\O_{f^{a_m}} = (\O_1, \O_2, \O_3)$$
		$$\O_{f^1} =   (-\O_1, \O_2, \O_3)$$
		
	\end{Cor}
	
	Corollary 1 means that during the saccadic cycle the oculomotor system can easily recalculate coordinates of axis vector w.r.t the primary frame from coordinates of it w.r.t the "moving" frame and vice versa at each switching point. In other words, the way of setting a saccadic cycle in terms of fixed axis vectors in Theorem 1 can be reformulated in terms of moving vectors in Theorem 2 and vice versa.


\end{document}